\documentclass[twocolumn,epjc3]{svjour3}

\RequirePackage{graphicx}

\usepackage{amsmath,amssymb,graphicx,color}
\usepackage[colorlinks=true,urlcolor=blue]{hyperref}
\usepackage{appendix}
\usepackage{enumitem}
\usepackage{url}

\journalname{Eur. Phys. J. A}

\begin{document}
\title{Topical issue on the intersection of low-energy nuclear structure and high-energy nuclear collisions}

 \author{Thomas Duguet\thanksref{addr1,addr2} \and Giuliano Giacalone\thanksref{addr5,addr3}  \and Vittorio Somà\thanksref{addr1} \and You Zhou\thanksref{addr4} }

\institute{IRFU, CEA, Université Paris-Saclay, 91191 Gif-sur-Yvette, France  \label{addr1} \and KU Leuven, Instituut voor Kern- en Stralingsfysica, 3001 Leuven, Belgium\label{addr2} \and Theoretical Physics Department, CERN, 1211 Geneva 23, Switzerland\label{addr5} \and Institut f\"ur Theoretische Physik, Universit\"at Heidelberg,
 Philosophenweg 16, 69120 Heidelberg, Germany\label{addr3}
 \and Niels Bohr Institute, University of Copenhagen, Jagtvej 155A, 2200 Copenhagen, Denmark\label{addr4}
            }

\date{}

\maketitle

Until very recently, the communities of high-energy heavy-ion physics and of low-energy nuclear structure physics have been largely disconnected. Heavy-ion physics relies on nuclear structure to provide the first step for the modeling of an ultrarelativistic collision between nuclei, with the goal of studying the properties of the quark-gluon plasma (QGP) that is subsequently created. Over the past two decades, nuclei used in high-energy collider experiments, mainly $^{208}$Pb and $^{197}$Au, have been assumed to present a structure plain enough to be captured in collision models by simple parametrizations of their charge densities, without invoking any notion of, e.g., neutron skin or nuclear deformation. However, with significant progress made in both experimental measurements and theory, our understanding of the hydrodynamic evolution in heavy-ion collisions is now precise enough to constrain also the initial condition of the created QGPs, prior to their hydrodynamic expansion. It is now understood that such an initial condition is sensitive to the details of the structural properties of the colliding nuclei \cite{Jia:2022ozr}, with nontrivial consequences on associated experimental observables.

This novel understanding has been made possible thanks to the relatively recent appearance of experimental data from multiple collision systems. Data from collisions of $^{238}$U nuclei \cite{STAR:2015mki,STAR:2024wgy,STAR:2025vbp}, released in 2015 from the BNL Relativistic Heavy Ion Collider (RHIC), can only be understood if the strong prolate deformation attributed to this nucleus through the confrontation of low-energy experimental data and nuclear structure models is included in the hydrodynamical simulation of the collision. Similarly, data from collisions of $^{129}$Xe nuclei collected in 2017 at the CERN Large Hadron Collider (LHC) \cite{ALICE:2018lao,CMS:2019cyz,ATLAS:2019dct} requires in addition the knowledge of the triaxial structure of this isotope \cite{ALICE:2021gxt,ATLAS:2022dov,ALICE:2024nqd}. The need for a precise knowledge of the structure of the colliding ions has been further emphasized in 2021 at the BNL RHIC with the release of experimental measurements on so-called isobar collisions, i.e., $^{96}$Ru+$^{96}$Ru and $^{96}$Zr+$^{96}$Zr collisions \cite{STAR:2021mii}. Ratios of bulk observables between such systems have been measured with unprecedented precision, and show clear manifestations of structural differences between $^{96}$Ru and $^{96}$Zr, including signatures of the larger neutron skin of $^{96}$Zr driven by its larger neutron number \cite{Giacalone:2025vxa}. All such results indicate that if one aims at a precise understanding of high-energy data, an equally precise modeling of the spatial structure of the colliding species is required.

In the state-of-the-art picture of high-energy nuclear collisions, the structure of nuclei impacts the formation of the QGP and its subsequent evolution because such experiments probe, on a collision-by-collision basis, nucleon configurations in nuclear ground states at the time of interaction, which is made possible by the ultra-short time lapse associated with the scattering process ($t\approx10^{-24}$ s). These nucleon configurations shape the geometry of the produced QGP. In the hydrodynamic model of heavy-ion collisions, such a geometry can, eventually, be reconstructed from experimental data. Having sensitivity to nucleon positions in individual nuclear configurations grants, then, access to many-body correlations in the collided ground states \cite{Duguet:2025hwi}. The technique by which many-body correlations become accessible in high-energy experiments is then akin to those employed in the study of highly-controllable quantum systems, such as cold atom gases, where the coordinates of individual constituents can be measured via imaging methods \cite{deJongh:2024pmo}. Remarkably, high-energy collider machines allow one to perform such an imaging of the atomic nucleus.

This opens a number of fundamental questions. The advances made at colliders provide us with a method to investigate whether emergent phenomena driven by QCD associated with atomic nuclei have a consistent phenomenology across vastly different energy scales. It is indeed not clear to what extent the Lorentz-boosted structure of nuclei probed in high-energy collisions is modified compared to that probed in low-energy experiments, and whether the semi-classical approximations employed in low-energy phenomenology remain valid in a high-energy environment. The quest for a quantitative understanding of the impact of nuclear structure properties (i.e., deformation, neutron skin, etc.) in high-energy collisions implies, in addition, a great synergy of collider data with the results of state-of-the-art nuclear structure calculations. This is especially timely regarding so-called \textit{ad initio} many-body calculations \cite{Hergert:2020bxy}, whose description of intermediate-mass doubly open-shell nuclei is expected to progress rapidly over the next few years, given that collisions involving oxygen-16 isotopes and other systems with $A \lesssim 100$ will be performed at the CERN LHC over the same period of time. Collider experiments thus represent a new tool to test and exploit the predictions of effective theories of low-energy QCD for nuclei. What are the implications of this established fact?

Making progress on these questions will have a broad impact on fundamental research in nuclear physics in the upcoming decades. Fostering and enhancing the collaboration between low-energy nuclear structure physicists and high-energy heavy-ion physicists to clarify issues and grasp physics opportunities at the intersection of these two areas is the primary aim of the proposed Topical Issue (TI). 

In its first part, the TI collects a set of introductory papers describing foundational principles underlying the hydrodynamic model of the QGP and the notion of deformation in nuclear structure from both experimental and theoretical viewpoints. These articles can serve as a comprehensive reference for physicists who aim to work at the intersection of nuclear structure and heavy-ion collisions in the future. The second part of the TI includes research articles presenting new advances in the field. A wide range of topics, involving the full breadth of ion species utilized so far in collider experiments, is covered, with the aim of advancing further this exciting research direction.

\bigskip
\noindent T. Duguet, G. Giacalone, V. Som\`a, Y. Zhou\\
\textit{Guest editors}


\section*{Table of contents}

\subsection*{1.\textbf{~Overview articles}}

\begin{enumerate}[label=(\alph*)]

    \item \textit{Measures of azimuthal anisotropy in high-energy collisions}\\J-Y. Ollitrault \cite{ollitrault}.
    
    \item \textit{Many-body correlations for nuclear physics across scales: from nuclei to quark-gluon plasmas to hadron distributions}\\G. Giacalone \cite{giacalone}.
    
    \item \textit{Effective theories for nuclei at high energies}\\O. G. Montero, S. Schlichting \cite{montero}.
    
    \item \textit{History of the concept of nuclear shape}\\D. Verney \cite{verney}.

\end{enumerate}

\subsection*{2.\textbf{~Research articles}}

\begin{enumerate}[label=(\alph*)]
  
  \item \textit{The shape of gold}\\B. Bally, M. Bender, G. Giacalone \cite{bally}.

  \item \textit{Impact of nuclear shape fluctuations in high-energy heavy ion collisions}\\A. Dimri,  S. Bhatta, J. Jia \cite{dimri}.

  \item \textit{Impact of nuclear deformation on collective flow observables in relativistic U+U collisions}\\N. Magdy \cite{magdy1}.

  \item \textit{Methods for systematic study of nuclear structure in high-energy collisions}\\M. Luzum, M. Hippert, J-Y. Ollitrault \cite{luzum}. 
  
  \item  \textit{Probe nuclear structure using the anisotropic flow at the Large Hadron Collider}\\Z. Lu, M. Zhao, X. Li, J. Jia, Y. Zhou \cite{lu}.

  \item \textit{Generic multi-particle transverse momentum correlations as a new tool for studying nuclear structure at the energy frontier}\\E. G. D. Nielsen, K. R\o mer, K. Gulbrandsen, Y. Zhou \cite{nielsen}.

  \item \textit{A study of nuclear structure of light nuclei at the electron-ion collider}\\N. Magdy, M. Hegazy, A. Rafaat, W. Li, A. Deshpande, A.M.H. Abdelhady, A.Y. Ellithi, R. A. Lacey, Z. Tu \cite{magdy2}.
    
  \item \textit{Prevailing triaxial shapes in atomic nuclei and a quantum theory of rotation of composite objects}\\T. Otsuka, Y. Tsunoda, N. Shimizu, Y. Utsuno, T. Abe, H. Ueno \cite{otsuka}.

  \item \textit{Angular structure of many-body correlations in atomic nuclei: From nuclear deformations to diffractive vector meson production in $\gamma A$ collisions}\\J-P. Blaizot, G. Giacalone \cite{blaizot}. 
      
\end{enumerate}

\end{document}